\documentclass[preprint,showpacs,amsmath,amssymb,aps,pra,floatfix,nofootinbib]{revtex4}
\usepackage{graphicx}
\usepackage{bm}
\newcommand{\be}{\begin{equation}}
\newcommand{\ee}{\end{equation}}
\newcommand{\bearr}{\begin{eqnarray}}
\newcommand{\eearr}{\end{eqnarray}}
\newcommand{\vrr}{\vec r}
\newcommand{\vrp}{\vec {r'}}

\begin{document}
\title{ Exact solutions of generalized Hubbard Hamiltonian for diamond vacancies }
\author{Mehdi Heidari Saani$^{*}$
, Mohammad Ali Vesaghi$^{\dag}$ 
and Keivan Esfarjani$^{\ddag}$} 
\affiliation{Department of Physics, Sharif University of Technology, Tehran,
P.O.Box: 11365-9161, Iran}
\begin{abstract}
A new formalism to calculate electronic states of vacancies in
diamond has been developed using many-body techniques.This model
is based on prevoius molecular models but does not use
configuration interaction and molecular orbital techniques. A
generalized Hubbard Hamiltonian which consists of all
electron-electron interaction terms is calculated on the atomic
orbital bases. Spatial symmetry $T_{d}$ and spin information of
system are included in the form of Hamiltonian, so the eigenstates
have automatically the correct spin and symmetry properties.
Optimizing two key parameters of the model that justifies already
reported semi-empirical values can predict accurate values of the
famous absorption lines in neutral and charged vacancies i.e.
$GR1$ and $ND1$. With these parameters the location of the low
lying $^{3}T_{1}$ state is $113$ mev above the ground state. In
addition to these levels whole of the energy states of the system
is predicted. Since the results are obtained without configuration
interaction the model can gives the exact contribution of
electronic configurations in the ground and excited states of the
neutral and charged vacancies. The new results of the model for
the ground and excited states of GR1 are reported.
\end{abstract}
\pacs{61.72.Bb, 61.72.Ji}\maketitle
\section{introduction}
Vacancy is one of the simplest and familiar intrinsic point
defects in semiconductors and its effects have attracted high
technological and theoretical interests.$^{1-3}$ A vacancy in
diamond in contrast to the situation in most semiconductor or
insulator crystals is stable in room temperature.$^{4}$ Since the
formation energy of vacancies in diamond is high ($6-7$ eV)
$^{5,6}$ they have been studied mainly in irradiated diamond.
Optical absorption studies are used for determination of
electronic excitation energies of vacancies$^{7,8}$ and EPR, ENDOR
experiments have been used for measuring the spin and spatial
symmetry of electronic states.$^{9-12}$ The most famous optical
absorption lines in irradiated diamond are $GR1$ (with zero phonon
line at $1.673$ eV)$^{13}$ and $ND1$ (with zero phonon line at
$3.149$ eV)$^{14}$. The former is attributed to a neutral
vacancy$^{15}$ and the latter to a negatively charged
one.$^{16}$\\In addition to these two famous lines many other
absorption lines have been reported such as: N3 ($2.985$
eV)$^{8}$, H3 ($2.463$  eV)$^{17}$, NV ($1.945$ eV)$^{18}$ which
have been attributed to vacancy-nitrogen$^{19,20}$, also GR2 to
GR8 lines$^{7}$ which have been attributed to the natural-charged
vacancy complexes.$^{21}$ Despite the simplicity of the system,
different models have been suggested to explain the properties of
these optical absorption lines. Generally there have been two main
approaches to this problem, The first group of approaches are
localized models$^{22-26}$ and the second ones are extended
models.$^{27-31}$ In the first group, the electrons of the broken
bonds which are in the tetrahedrally symmetric local potential of
the vacant site are considered as an isolated molecule (many body
approaches). In the second group i.e. extended models, the effect
of lattice is more important on the vacancy and it is considered
as a missing atom in a supercell with few hundreds of atoms
(single particle approaches). Similarity between these two
categories is their group theoretical aspects for accounting
symmetries of the system while their main difference is in the
importance of e-e correlation effects. Lannoo and Bourgoin
$^{32}$have developed a model with four parameters which
briliantly explaines the validity range of the single and many
particle approaches for this problem. According to Messmer and
Watkins$^{27}$ a successful model for describing vacancy energy
states should explain the following points: The position of the
defect energy levels relative to the valence band edge, the
electronic wavefunction of defect electrons for comparing with
experimental data, lattice relaxation and distortion in
neighborhood of vacant site and it must be the basis of a
practical computational scheme. The localized approaches have
relatively successful answers to these questions.
\\The first suggested model in this regard is the famous model of
the Coulson and Kearsly$^{33}$. Their model suggests a
computational scheme that considers e-e interaction in the system
accurately. They used group theoretical methods to  manually
construct symmetry and spin adapted wavefunctions for including
spin and symmetry considerations of vacancies. By applying
configuration interaction technique to the calculation results,
they predicted qualitatively good results for electronic states of
the vacancies, their model used two semi-empirical parameters to
obtain satisfactory result for energy of the $GR1$ transition. By
using these semi-empirical parameters their predictions for $ND1$
transiton and also the posision of the low lying $^{3}T_{1}$ state
disagrees with avaiable experimental data. Up to now this model
has been the most successful computational scheme for describing
physical properties of the optical absorption lines of vacancies
in diamond,$^{3,21}$ and the next suggested localized
models$^{21-26}$ are based on the framework of this model.\\
Localized models, unlike the nowadays commonly used density
functional theory (DFT) methods which pay more attention on the
ground state information, can also give the excitation energies of
the vacancies beyond the one electron approximation.
\\For improving the quantitative agreement between theory and
experiment we introduce a new approach for localized models. This
approach is based on a generalized Hubbard Hamiltonian for
electrons of the vacancy system with atomic orbital bases and uses
computer adaptable many body techniques instead of usually used
molecular orbital method (fourth item of messmer and watkins). In
this paper at first we review the computational scheme of the
present model and discuss the advantages of the new notation which
is used for this problem. Then we apply this new scheme to solve
generalized form of Hubbard Hamiltonian for neutral and charged
vacancy systems. For evaluation of solutions of this Hamiltonian
we start with parameters which were already reported. Our using
value for the two semi-empirical parameters slightly differs from
already rpoeted values but other six parameters are as the same as
the theory calculation. We report the effect of the hopping
parameter on the energy spectrum of this Hamiltonian by varying this parameter.\\
The model predictions for $GR1$ and $ND1$ transitions with
justified parameters and comparison the results with experimental
data will be discussed. The location of the low lying $^{3}T_{1}$
state is predicted and we disscuss its comparision with other
theoretical reports and experimental evidences. The results of our
model can justify the semiempirical values that already were
reported.$^{33,41}$ We also will show that, this model is
independent of familiar configuration interaction technique and
the electronic configurations mixing of each state comes out in a
natural manner from the formalism. The model predicts whole of the
exact eigenstates and enery levels of the systems with the
justified semi-empirical values. Finally the new information
concerning electronic configurations of the ground and excited
states of $GR1$ transition are reported which can be used by other
approaches to this problem.
\\In our calculation the effects of lattice relaxation were
ignored, which according to recent theoretical works are
small.$^{34-36}$
\section{Computational model}
The present computational model takes into account electron
correlation effects accurately and gives the correct spin and
spatial symmetry of eigenstates. It is also suitable for small
atomic cluster systems or molecules provided the hopping,
Coulombic and exchange integrals are known. In the localized model
framework, it is assumed that physical properties of neutral
(charged) vacancy depend on four (five) electrons of adjacent
dangling bounds. The exact Hamiltonian of such a system can be
written as Eq. (1). A more simplified form of this Hamiltonian
recently is applied to carbon nanotube systems $^{36,37}$ .
\begin{eqnarray}
H=\sum_{ij\sigma}t_{ij}c_{i\sigma}^{\dag}c_{j\sigma}+\frac{1}{2}\sum_{ijlm\sigma\sigma^{`}}V_{ijlm}c_{i\sigma}^{\dag}c_{j\sigma^{`}}^{\dag}c_{m\sigma^{`}}c_{l\sigma}
\end{eqnarray}
Where $i,j,l,m$ are atomic sites indices which go from $1$ to $4$
, $\sigma$ is spin index which is $-1/2$ or $+1/2$ ,
$c^{\dag}_{i\sigma}$ and $c_{i\sigma}$ are operators which create
and annihilate electrons with specified spin $\sigma$ on site $i$.
 $t_{ij}$ and $V_{ijlm}$ parameters are hopping and $e-e$
interaction overlap integrals which are defined as follows:
\begin{eqnarray}
 t_{ij}= \int\int d\vrr \psi_{i}^*(\vrr)
(\frac{-\hbar^{2}}{2m}\nabla^{2}+V(r))
 \psi_{j}(\vrr)
\end{eqnarray}
\begin{eqnarray}
 V_{ijlm}  =
 \int\int d\vrr d\vrp~\psi_i^*(\vrr)\psi_j^*(\vrr)\frac{1}{|r-r'|}
 \psi_l(\vrp)\psi_m(\vrp)
\end{eqnarray}
in the above integrals $\psi_{i}$ is single electronic wave
function in a specified site. The advantage of using second
quantized form of Hamiltonian for this problem is its capability
to include the spatial symmetry and spin information of the system
in the form of Hamiltonian, hence the eigenstates have
automatically the desired spin and symmetry properties. Also the
effect of the changing each parameter directly can be observed on
the electronic energy levels and also wave functions. This new
approach uses atomic orbital bases to solve the Hamiltonian. In
the previous localized models it was common as a start point, to
manually construct symmetry and spin adapted molecular wave
functions as correct basis for accounting symmetry considerations
of the system then these basis should be used in evaluating
Hamiltonian matrix elements. This process is a laborious task and
can not be incorporated in a computational software easily.
\\Unlike previous methods, present computational scheme needs
significantly less effort and is straightforward in converting to
computer software language. The starting point of the model is
constructing appropriate many-body basis needed for solving the
Hamiltonian.\\ All of the possible distribution of electrons in
the four adjacent atomic orbital make a set of complete basis in
the configuration space. In contrast to previous models that have
used molecular orbital for such a system we have developed atomic
orbital bases by taking into account $z$ component of the spin
degree of freedom with occupation condition of each orbital.
Following notation can be used to represent each configuration
state. Ignoring the spin degree of freedom simply reduces the
states to the occupation number or the Fock space representation.
Fock space representation is commonly used in modern molecular
calculations.$^{39}$
\begin{equation}
  | \Psi_{i}>=|a_{i0},a_{i1},a_{i2},...,a_{i8}>
\end{equation}
Parameters $a_{i1}$ to $a_{i8}$ are -1 or +1 (for spin down and up
on the site $i$ respectively) which show occupation condition of
each state. In fact these coefficients not only show extension of
electronic wave function on each neighboring site of vacancy but
also include spin information of the system. $a_{i0}$ is the sum
of all $ a_{i}$'s i.e. the total spin of each $\Psi_{i}$ along $z$
axis. We will show the importance of this quantum number in our
computation scheme later. Other capability of this notation is
estimation of the maximum dimension of configuration space which
one needs for calculating the Hamiltonian for $V^{0}$ $(V^{-})$.
In the $V^{0}$ $(V^{-})$ system there are four (five) electrons
respectively in four orbital, so the possible configurations of
these electrons in the $8$ accessible states ($2$ spin and $4$
space or site degree of freedom) according to statistical
combination rules are: \be\label{2-3} C^{4}_{8}=70 \hspace{2.5cm}
(V^{0}) \ee \be\label{2-4} C^{5}_{8}=56 \hspace{2.5cm} (V^{-})\ee
The fractions of $70$ $(56)$ configuration states in $V^{0}$
$(V^{-})$ system which belong to each $S_{z}$ block are summarized
in Table I. This counting method is in contrast to the old
molecular orbital approaches which one needs to allow construction
of symmetry and spin
adapted basis by hand and then counting them. \\
After constructing the set of complete basis, we will calculate
two form of Hamiltonian in this space. The first one is
approximate form of Eq. (1), i.e. a simple extension of Hubbard
model and the second is the exact form of the Hamiltonian (Eq.
(1)) which we call it generalized Hubbard Hamiltonian.
\begin{widetext}
\begingroup
\squeezetable
\begin{table}[t]
\begin{tabular}{ c c c c c c c c c c c c c c c c c c}
\hline \hline
&\multicolumn{10}{c}{$ V^{0}$}& \multicolumn{6}{c}{$ V^{-} $}\\
\hline $S^{tot}_{z}$&&&+2&+1&0&-1&-2&&&&+3/2&+1/2&-1/2&-3/2&&\\
$n_{up}$&&&4&3&2&1&0&&&&4&3&2&1&&\\
$n_{\Psi}$&&&1&16&36&16&1&&&&4&24&24&4&&\\
\hline \hline
\end{tabular}
\caption{In this table, $S^{tot}_z$ is total spin in $z$ axis
direction, $n_{up}$ is number of electrons with spin up and
$n_{\Psi}$ is number of states with specific $S_z$.}
\end{table}
\endgroup
\end{widetext}
\section{Generalized Hubbard Hamiltonian}
The main goal of the Hubbard Hamiltonian$^{40}$ is resuming
atomistic nature of the solid besides the free electron gas
theory. This model assumes that the most important part of $e-e$
interaction terms is on site term on the Columbic parts of
Hamiltonian. The conventional form of this Hamiltonian consists of
two parts, hopping term or $t$ term and on site term or $U$ term
which is $V_{iiii}$ term of Eq. (3).
\begin{equation}
H=\sum_{ij\sigma}t_{ij}c_{i\sigma}^{\dag}c_{j\sigma}+\sum_{i}U_{i}n_{i}\uparrow
  n_{i}\downarrow
\end{equation}
Famous Hubbard model with two parameters $t$ and $U$ is not
sufficient for solving the vacancy problem, the results of our
model with these two parameters are not satisfactory. This result
is physically expecting since assuming only two parameters of the
Hubbard model for this system means that the electrons interact
with each other only when they are at the same site of at the same
atom of the vacancy and this eliminates the interaction of the
electrons when they are in the different atomic site of the
vacancy. Therefore we started with extended Hubbard Hamiltonian
with one extra parameter $V$ which is $V_{ijij}$ term of Eq. (3).
This Hamiltonian is equivalent to the Lannoo model$^{32}$ where
the $U$ parameter of their model is called $V$ in our model. This
means that the on site and two site direct overlap integrals are
the most dominant ones and the rest parameters are negligible
(Eq.(8)).
\begin{widetext}
\begin{equation}
  H=\sum_{ij\sigma}t_{ij}c_{i\sigma}^{\dag}c_{j\sigma}+
\sum_{i} U_{i}n_{i}\uparrow n_{i}\downarrow
+\frac{1}{2}\sum_{i\neq j \sigma \sigma^{`}}
V_{ij}n_{i\sigma}n_{j\sigma^{`}}
\end{equation}
\end{widetext}
Since the atomic sites in vacancy are from the vertex of a
symmetrical tetrahedral, the distance between each two of them is
the same, and hopping $(t_{ij})$, on site $(U_{i})$ and two site
$(V_{ij})$ parameters of Hamiltonian are all independent of $i,j$
indices and can be put outside of the sum Eq. (8).
\begin{equation}
  H=t\sum_{ij\sigma}c_{i\sigma}^{\dag}c_{j\sigma}+U\sum_{i}n_{i}\uparrow
  n_{i}\downarrow +\frac{1}{2}V\sum_{i\neq j \sigma \sigma^{`}} n_{i\sigma}n_{j\sigma^{`}}
\end{equation}
As discussed in sec. II, for $ V^{0} $ $(V^{-})$ system the size
of the set which includes complete basis for solving this
Hamiltonian is $70$ $(56)$. As a result solving Hamiltonian in
this set results in a $ 70\times70$ ($ 56\times56 $) Hamiltonian
matrix for $V^{0}$ ($V^{-}$) respectively. As the starting point
we used numerical values of Hamiltonian parameters $t$, $U$ and
$V$ which were reported by semi-empirical and theoretical works on
diamond vacancies.$^{33,41}$\\
This formalism can investigate the effect of variation of each
parameter on the energy spectrum  directly. By this means we found
that variation of parameter $t$ has not any effect on the $GR1$
transition energy. But variation of parameters $U$ and $V$ changes
the transition energy of $GR1$. These variations do not affect the
sequence of the levels. The results of calculation for $V^{0}$
show that for a wide variation range of variables, $t$ greater
than $-12.5$ eV, $U$ greater than $8.2$ eV and $V$ greater than
$4.0$ eV, the ground state of the system has $^{1}E$ symmetry
(double degenerate spin-less) and we can obtain appropriate
transition to an excited state $^{1}T_{2}$ (triple degenerate
spin-less). With our model we were able to obtain the experimental
value of $1.673$ eV for the GR1 transition.
 \\For $V^{-}$ system, variation of $t$ has a
dramatic effect on $ND1$ transition energy, but does not alter the
ground and excited states sequence. Unlike this, $U$ and $V$
parameters which have only small effect on the values of the
transition energies of the system. For the same range of variation
of parameters we had a transition from $^{4}A_{2}$ ground state to
a $^{4}T_{1}$ excited state ($ND1$ transition) but By the same
parameters which we obtained the transition energy of $GR1$
($1.673$ eV), there is a wide gap between the
ground and excited states of $ND1$ transition (about $30$ eV).\\
Only with parameter $t$ close to $-1$  was the result close to the
experimental value. This high value for $t$ seems
unreasonable for the vacancy problem and is very far from reported range.$^{33,41}$\\
These results of the extended Hubbard Hamiltonian are in agreement
with the results of the Lannoo and Bourgoin $^{32}$ model which
has simplified the full Hamiltonian in the first quantization form
by four parameters. They have used symmetrical molecular orbital
base and also configuration interaction similar to Coulson and
Kearsly model. The benefith of the reduction of the number of
Hamiltonian parameter in their model was in this point that the
validity range of single and many particle approches could be
investigated. They investigated the results of their model in a
full range of parameters and also used the semi-empirical
parameters of the Coulson and Kearsly model. They successfully
explained where the many body model works and where the single
particle approaches can be valid. Lowther$^{21}$ applied a similar
model to the Lannoo model to explain some related optical
phenomenon of the vacancies in diamond. Mainwood and
Stonham$^{26}$ applied Lannoo model directly to the vacancies
electronic states in diamond.
 It seems that lack of the quantitative agreement
with observed $ND1$ transition energy ($3.15$ eV) from $^{4}A_{2}$
ground state to a $^{4}T_{1}$ excited state in $V^{-}$ system is
due to elimination of exchange terms in Hamiltonian of vacancy
system. Since calculation results of $V^{-}$ system show weaker
agreement with the experiment than those for $V^{0}$, we can
deduce that the role of the exchange terms in charged electron
vacancy which has higher electron density is more important than
that the four electron (neutral) one. Quantitative failure of the
results of this simple extension of Hubbard Hamiltonian reveals
that for vacancy system $e-e$ correlation effects are important
and can not be neglected in contrast to the other cases which we
are able to apply the Hubbard model. This point also was
emphasized by previous models which have attempted diamond vacancy
problem.$^{21-26,33,41}$ Comparing numerical values of overlap
integrals of Hubbard model$^{40}$ and diamond vacancy
problem$^{33,41}$ (Table II) we find that the values of exchange
integrals in this problem are comparable to the values of direct
integrals so their elimination are not reasonable.
\begin{widetext}
\begingroup
\squeezetable
\begin{table}[t]
\begin{center}
\begin{tabular}{c c c c c c c}
\hline \hline&{$V_{iiii}$}& {$V_{ijij}$}&
{$ V_{iiij}$}& {$ V_{ijik}$}& {$V_{iijj}$}\\
\hline Hubbard model$^{40}$\hspace{0.8cm}(eV)&20&2-3&0.5&0.1&0.025\\
\hline Diamond vacancies$^{33,41}$ (eV)&12.855&7.851&1.30&0.249&0.419\\
\hline \hline
\end{tabular}
\caption{List of estimated Hamiltonian parameters in Hubbard model
and diamond vacancies problem.}
\end{center}
\end{table}
\endgroup
\end{widetext}
This leads us to keep all of the, $t$, $U$ and $V$ terms besides
exchange terms in four (neutral) and five (charged) electronic
Hamiltonian of the system similar to Coulson and Kearsly
model.$^{33,41}$ We will call it generalized Hubbard Hamiltonian
(Eq. (1)). Although this significantly increases the volume and
complexity of the calculation, but since we take into account
exact $e-e$ correlation effects, it makes our model more
realistic. \\We return to Hamiltonian form of Eq. (1). Similar
symmetrical argument is valid for a system with $T_{d}$ symmetry
and we can put hopping parameter $(t)$ out of the sum (Eq. (10)).
\begin{equation}
  H=t\sum_{ij\sigma}c_{i\sigma}^{\dag}c_{j\sigma}+\frac{1}{2}\sum_{ijlm \sigma \sigma^{`}}
  V_{ijlm}c_{i\sigma}^{\dag}c_{j\sigma^{`}}^{\dag}c_{m\sigma^{`}}c_{l\sigma}
\end{equation}
Symmetry of defect molecule, reduces the number of the independent
parameters $V_{ijlm}$, from $4^4$ $(256)$  components to only $7$
independent ones, hence symmetry considerations can be included in
the generalized form of Hamiltonian with a total of $8$
parameters. These consist of the hopping $t$, two direct Coulombic
 integrals $(U,V)$ and five exchange $(X1,...,X5)$ integrals as follows:
\begin{eqnarray*}
 U=\langle ii|\frac{1}{r}|ii\rangle\\
 V=\langle ij|\frac{1}{r}|ij\rangle\\
 X1=\langle ij|\frac{1}{r}|ji\rangle\\
 X2=\langle ii|\frac{1}{r}|ij\rangle\\
 X3=\langle ij|\frac{1}{r}|ik\rangle\\
 X4=\langle ii|\frac{1}{r}|jk\rangle\\
 X5=\langle ij|\frac{1}{r}|kl\rangle\\
\end{eqnarray*}
Since the Hamiltonian is spin independent, useful conserving
quantum numbers
\begin{equation}\label{a}
  [H,S^{2}]=[H,S_{z}]=0
\end{equation}
are total spin of four electron system $(S^{2})$ and the $z$
component of total spin $(S_{z})$. The important computational
point about Eq. (11) is that it should hold for each arbitrary set
of parameters of Eq. (10).\\ By these conservation rules we can
easily decrease dimension of the nonzero block of Hamiltonian
matrix by using conservation of total spin ($S^{2}$) and $z$
component of the total spin ($S_{z}$). Therefore the volume of
calculation significantly reduces and matrix element evaluation
only confines in subspaces where wave functions have the same
values of $S_{z}$. As a result, the maximum dimension of nonzero
matrix block in the case of $V^{0}$ ($V^{-}$) systems becomes $36$
$(24)$ respectively. The final step in solving the Hamiltonian of
the $V^{0}$ ($V^{-}$) for obtaining the resultant eigenstates with
definite $S_{z}$ and $S^{2}$ values is the transformation of
Hamiltonian matrix from $S_{z}$ to ($S^{2},S_{z}$) basis.\\
We have used the parameters which were calculated by Coulson and
Larkins$^{41}$  to investigate energy spectrum and eigenstates of
the system. These parameters have been calculated
using Slater and Clementi type functions as atomic orbitals. \\
The parameter calculation is independent from the Hamiltonian
solution formalism. Since the parameters are calculated two times
by the Coulson $^{33,41}$ during a long period, other models
similar to us have not attempted to re-estimate their values and
all of them have used the results of the Coulson estimation for
the parameters. $^{21,22,26,32}$The calculation of parameters were
based on symmetric and antisymmetric molecular orbital basis ($
a,t_{x},t_{y},t_{z}$)which were combinations of $a,b,c,d$ atomic
orbitals. These atomic orbitals belong to the four nearest
neighbor atoms of vacant site and are oriented inward to its
center, however in the generalized Hubbrad Hamiltonian Eq. (1),
parameters are calculated directly by using overlap integrals of
these $a,b,c,d$ atomic orbitals. These overlap integrals can be
extracted from the older symmetric and anti-symmetric ones by
changing the calculation basis. Related calculations have been
performed and the resultant parameters are used in the
computational scheme.
\section{Results and discussion}
The two key parameters of this model are $t$ and $U$. These two
have been evaluated semi-empirically in the previous models and
were starting point in calculation of the other
parameters.$^{33,41}$ The reported values for hopping parameter
$t$ cover a wide range of variations, for example both $-7.13$ eV
and $-16.34$ eV values at the same time are reported for this
parameter.$^{33,41}$ Concerning parameter $U$, the semi-empirical
value is estimated to be $13.29$ eV and after revising $12.85$ eV
while the theoretically calculated value is $19$ eV .$^{33,41}$
\begin{widetext}
\begingroup
\squeezetable
\begin{table}[t]
\begin{center}
\begin{tabular}{c c c c c c c c c c}
\hline\hline&{$ t $}& {$ U $}& {$ V $}& {$ X1 $}& {$ X2 $}& {$ X3$}& {$ X4 $}& {$ X5 $}\\
\hline Present model\hspace{0.4cm}(eV)&-7.747 &13.58&7.851&0.419&2.18&0.249&1.30&0.215\\
 Ref. 33,41\hspace{0.9cm}(eV)&-8.12&12.855&7.851&0.419&2.18&0.249&1.30&0.215\\
 Ref. 41   \hspace{1.2cm}(eV)&-7.13&12.855&7.851&0.641&2.663&0.351&1.563&0.318\\
\hline \hline
\end{tabular}
\caption{List of eight parameters that are needed to evaluate
energy states of vacancy. They are estimated by this work and
other different calculations.}
\end{center}
\end{table}
\endgroup
\end{widetext}
\begin{center}
\begin{figure}[b]
\includegraphics[width=5cm,height=7cm,angle=270]{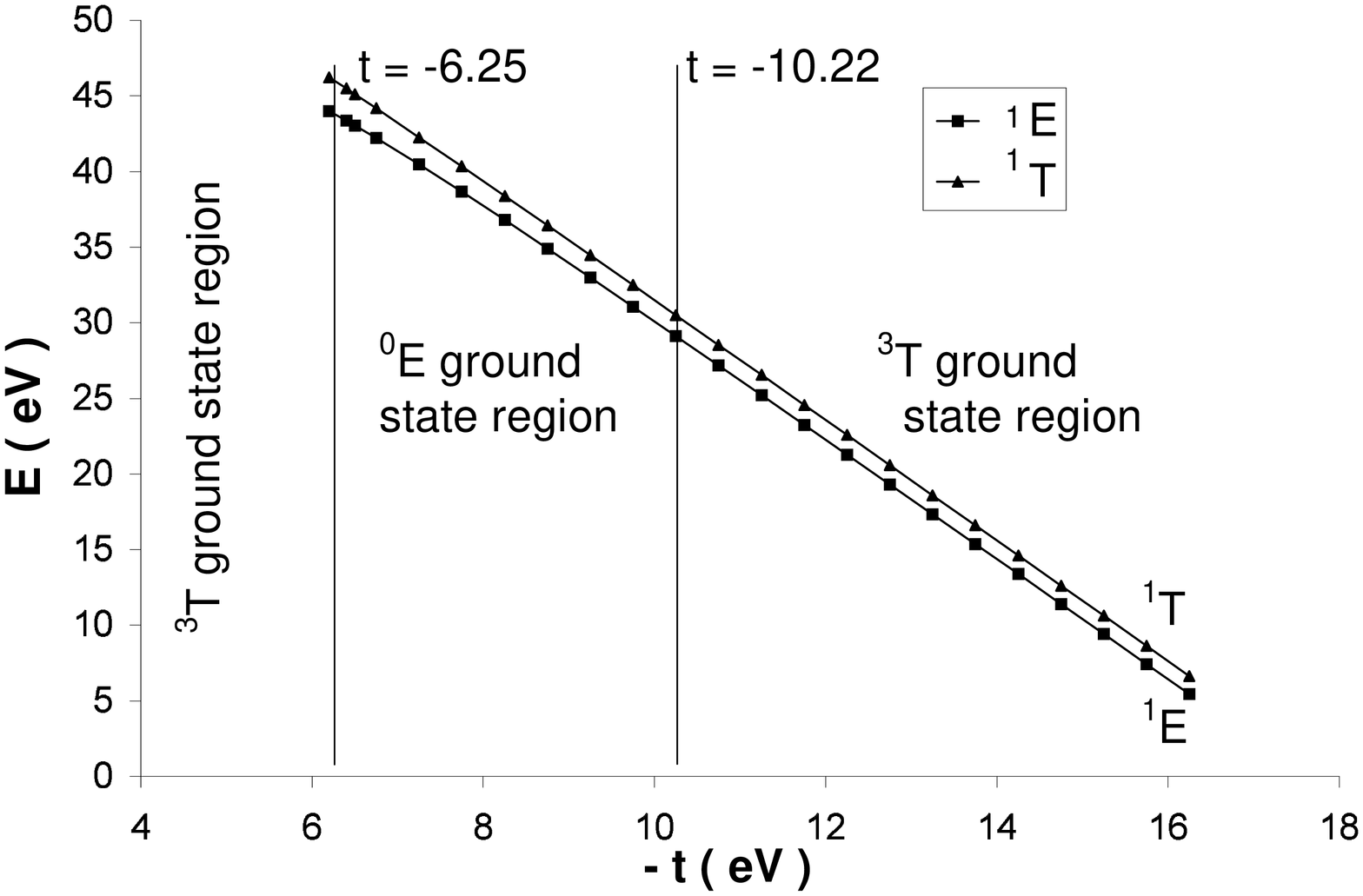}
\caption{Variation of ground $^{1}E$ and excited $^{1}T_{2}$
states of $GR1$ transition, with respect to hopping parameter $t$.
The $t=-6.25$ eV, $-10.22$ eV are changing point of the ground
state with $^{3}T_{1}$ state.}
\end{figure}
\end{center}
In evaluation of the $t$ parameter despite of other $7$ Coulombic
parameters, one not only need the information of the spatial
distribution of the wave functions but also the form of the
psedupotential of carbon atoms is needed, so the uncertainty in
the value of this is more than other parameters. In the present
model the variation of the energy spectrum with $t$ has been
investigated and results are shown in Fig. 1. The values of the
other parameters are according to table III. According to this
figure, for obtaining correct results for ground state of $V^{0}$
i.e. $^{1}E$, parameter $t$ should be confined in the following
range.
\begin{equation}
-6.25 eV\geq t \geq-10.22 eV
\end{equation}
Choosing $t$ out of this range converts the sequence of two lowest
states i.e. $^{1}E$ and $^{3}T_{1}$, and gives wrong ground state
($^{3}T_{1}$) for the system. Increasing value of the $t$ will
lower the experimentally observed spin quintet $^{5}A_{2}$
state.$^{11}$ It is interesting to note that above the upper limit
of $t$, $^{5}A_{2}$ competes with $^{3}T_{1}$ to be the ground
state and at $t$ value equals to $-5.96$ eV, $^{5}A_{2}$ becomes
ground state of $V^{0}$ system. It seems that by increasing $t$ we
reach an area where the Hund's rule can be applied (according to
this rule $^{5}A_{2}$ must be the ground state). This might be due
to the point that $t$ is the indicator of the looseness of the
electrons from the nuclei so by
increasing $t$ we can consider the electrons as more delocalized electrons.\\
For charged vacancy case, the calculation results are shown in
Fig.2. From this
\begin{center}
\begin{figure}[t]
\includegraphics[width=5cm,height=7cm,angle=270]{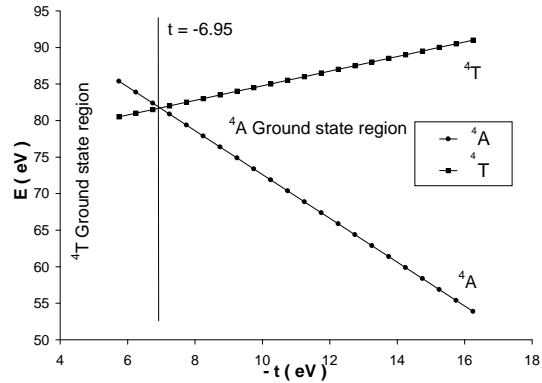}
\caption{ Variation of ground $^{4}A_{2}$ and excited $^{4}T_{1}$
states of $ND1$ with respect to hopping parameters $t$. The
$t=-6.95$ eV is changing point of the ground state with the
excited state.}
\end{figure}
\end{center}
figure we conclude that for obtaining $^{4}A_{2}$ as ground state
for the system, parameter $t$ should be restricted as:
\begin{equation}
-6.95 eV\geq t \geq-16.34 eV
\end{equation}
The lower limit is the negative of the ionization energy of
electron in C-C bound of diamond. From this figure we can find
that increasing $t$ over upper limit converts the ground state to
$^{4}T_{1}$ which disagrees with the experiment. \\
By comparing the results of the calculation for both $V^{0}$ and
$V^{-}$ we can put new boundaries to the single particle hopping
parameter $t$ as follows:
\begin{equation}
-6.95 eV\geq t \geq-10.22 eV
\end{equation}
This restriction on $t$ parameter, which is the most uncertain
parameter in the vacancy electronic states problem, is tighter
than already reported  range ( $-7.13$ eV to $-16.14$
eV).$^{33,41}$ Variation of $t$ in this allowed range does not
have significant effect on $GR1$ transition energy. This is in
agreement with other theoretical works.$^{33,41}$ However this
variation has a dramatic effect on $ND1$ transition energy.\\
Numerical optimization of parameter $t$ in this range and also
parameter $U$ gives $-7.747$ eV and $13.58$ eV for $t$ and $U$
respectively. These values are very close to the semi-empirically
derived values by Coulson and Larkins (the difference between our
optimized value by semi-empirical values in this two parameters
are less than 8 and 5 percent respectively). $^{41}$ Other six
parameters are as the same as the theoretically calculated values
by them using Slater type functions $^{41}$. Calculation methods
of these parameters are completely independent from the
Hamiltonian solution formalism so the previous models
$^{21,22,26,32}$ similar to us have used the parameters which are
reported semi-empirically and theoretically by Coulson and Kearsly
$^{33}$and Coulson and Larkins.$^{41}$ These parameters are
calculated by them during a 17 year period and with the Slater
type functions so it seems that the parameters are well justified.
The difference between our suggested parameters with two
semi-empirical parameters of their model is very less than the
wide range of variation of these parameters in their model.(
parameter t from $-7.13$ eV to $-16.34$ eV and parameter U which
smi-empirically is $12.855$ eV and theoretically is $19$ eV)\\
With these justified values the present model can simultaneously
obtain the exact transition energies of  the $GR1$ ($1.67$ eV) and
$ND1$ ($3.15$ eV) for neutral and charged vacancy system.
\\The set of our parameters and parameters that were previously
computed based on Slater and Clementi type functions$^{33,41,42}$
are listed in Table (III). The model predicts the wrong ground
state with parameters which are based on Clementi type functions,
i.e. $^{3}T_{1}$ instead of the experimentally observed $^{1}E$
state for $V^{0}$. This is similar to the Coulson and Larkins's
report. $^{41}$ However parameters which are based on the Slater
type functions predict correct spin and symmetry for ground and
excited states of $GR1$ and $ND1$ absorption lines but the values
of the transition energies are not satisfactory. The results of
the present model calculation are summarized in Fig. 3,4.\\ For
neutral vacancy as it is shown in Fig. 3, there is a transition
from double degenerate spinless ground state to a triple
\begin{center}
\begin{figure}[h]
\includegraphics[width=5cm,height=7cm,angle=270]{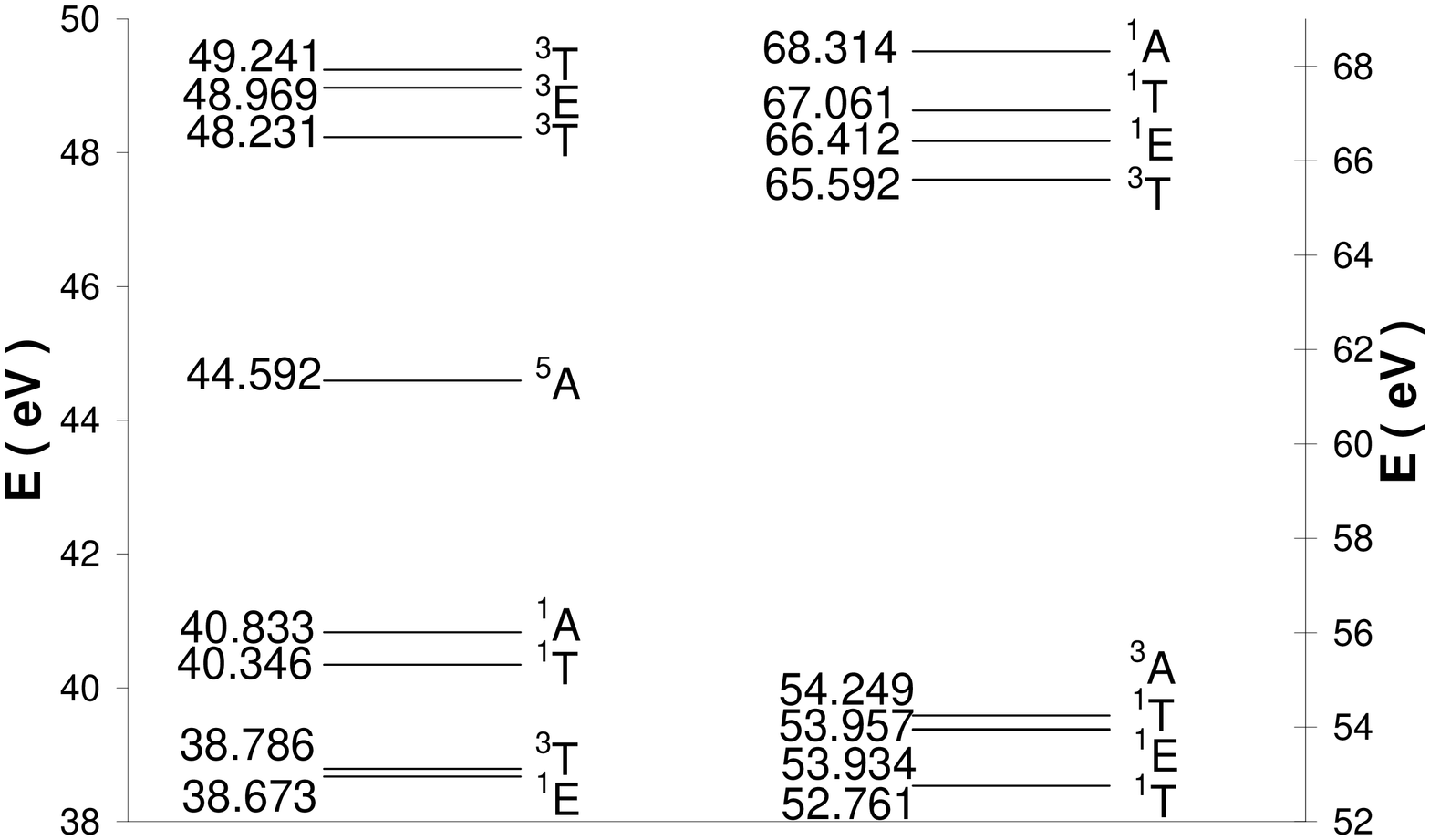}
\caption{ Neutral vacancy energy levels
 with, $t=-7.747$ eV, $U=13.58$ eV (present parameters)
 gives $1.673$ eV for $^{1}E$ to $^{1}T_{2}$ transition.
 The right-hand side levels are the continuation of the left-hand side levels.}
\end{figure}
\end{center}
degenerate spinless excited state
$(^{1}E\longrightarrow^{1}T_{2})$ with a transition energy equal
to $1.67$ eV.\\We also see that there is another level
$^{3}T_{1}$, very close to $^{1}E$ ground state with a very little
energy difference i.e. about $113$ meV above it. This value agrees
with the recent experimental expectations which predict it will be
more than $100$ meV .$^{43}$ However the results of Coulson models
gives $40$ meV which disagree with the experiment.$^{33,41}$ This
very close energy level to ground state also has been used to
explain the behavior of $R2$ center absorption line in diamond
where it was attributed to the strongly perturbed vacancy.$^{42}$
By this assumption the energy difference of this level with the
ground state has been predicted to be between $40$ meV and $200$
meV.$^{42}$ Mainwood and Stonham $^{26}$ have used the simple
model of Lannoo $^{31}$ and successfully explain many physical
features of the vacancy. Lannoo for simplifying the problem do not
enters the exchange parameters into energy levels and use only two
parameters for describing energy levels. This is unlike to Coulson
$^{32,40}$ and our model which reserve whole of the e-e terms.
Besides explaining many physical feature of the vacancy they gave
some estimation for the position of this level. Using the model of
Lannoo they positioned $^{3}T_{1}$ at $200$ meV above the $^{1}E$,
Although this value is as the same magnitude as the error in the
positions of the levels in their calculations.$^{26}$\\
Our reported location for $^{3}T_{1}$ arises in a natural manner
similar to $GR1$ and $ND1$ after evaluation of the exact
Hamiltonian of the system with the justified parameters. The error
which can be attributed to this value in our calculation is the
difference between our parameters and semi-empirical parameters of
the Coulson. This error is less than $8$ percent which is
significantly less than already reported error for the position of
this level in similar calculations.$^{26}$ In obtaining this value
we have not assumed that the vacancy is highly perturbed as it
already was reported.$^{42}$ This value can help EPR experimental
investigations that expect to observe this state in temperatures
higher than $300$ K.$^{43}$\\
\begin{center}
\begin{figure}[h]
\includegraphics[width=5cm,height=7cm,angle=270]{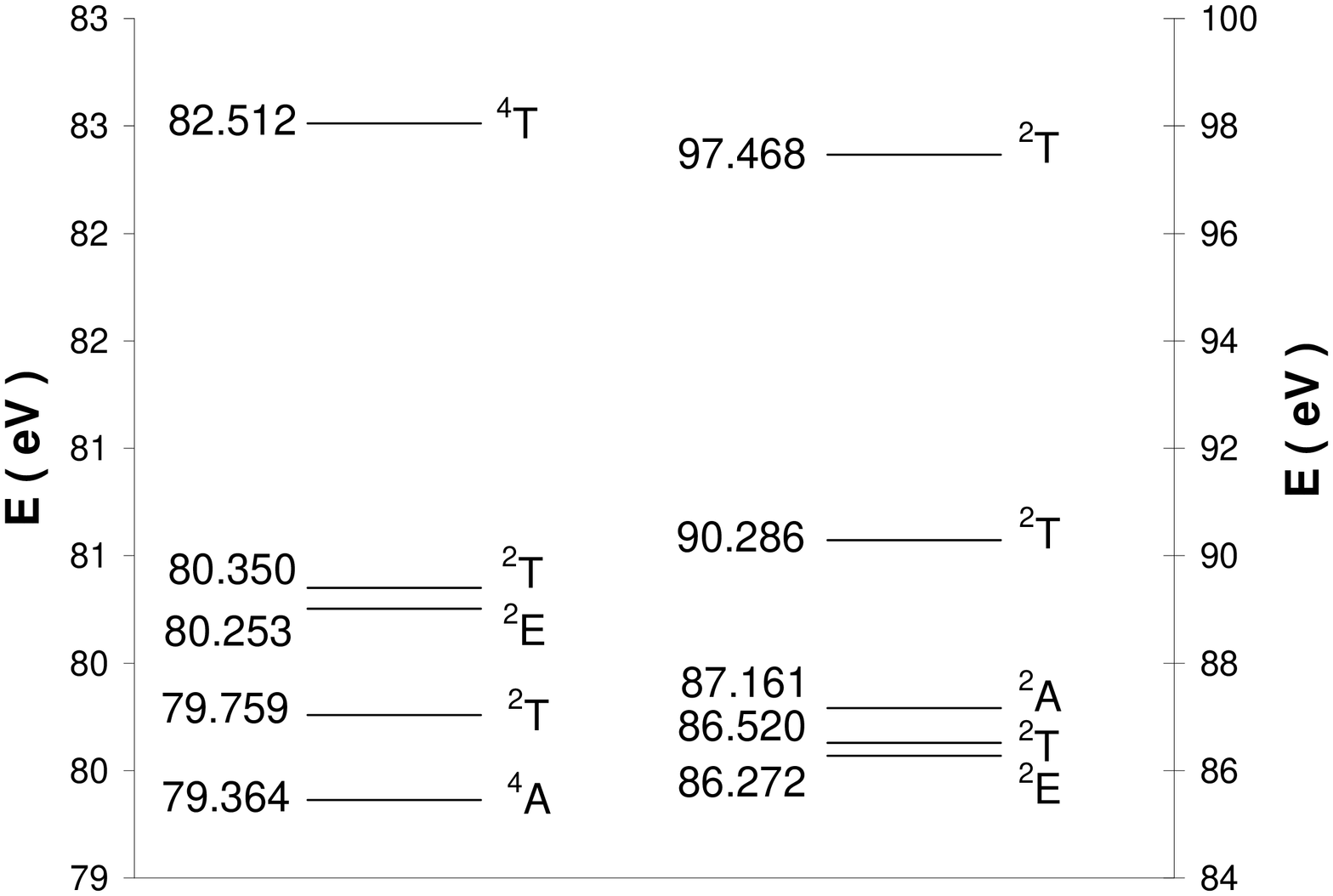}
\caption{ Negative vacancy energy levels with, $t=-7.747$ eV,
$U=13.58$ eV (present parameters) gives $3.159$ eV for $^{4}A_{2}$
 to $^{4}T_{1}$ transition. The right-hand side levels are the continuation of the left-hand side levels.}
\end{figure}
\end{center}
In the charged vacancy system, as it is shown in Fig. 4, the model
predicts a transition from a non-degenerate ground state to a
triple degenerate excited state with a spin equal to $3/2$
$(^{4}A_{2}\longrightarrow^{4}T_{1})$. The energy difference is
$3.15$ eV which is the famous observed $ND1$ absorption line.
Previous models that evaluate e-e interaction exactly can not
obtain this value by any set of parameters. $^{33,41}$
\\In summary as it has been mentioned by other authors $^{32,26}$
the previous models that evaluate e-e interaction exactly
$^{33,41}$ can only explain GR1 transition energy quantitatively.
They could not predict $ND1$ transition energy and also they have
predicted the experimentally wrong energy for the low-lying
$^{3}T_{1}$ level. These unsatisfactory quantitative results are
obtained not only by the semi-empirical parameters but also in a
wide variation range of these two semi-empirical
parameters.$^{33,41}$ This variation ranges is very wider than the
difference of our optimized values with semi-empirically reported
ones for $t$ and $U$. Other six Coulombic parameters in our model
and the previous ones are evaluated by theory.$^{33,41}$ Our model
can simultaneously give accurate transition energies for both of
 the $GR1$ and $ND1$ transitions besides the experimentally acceptable
energy for the low-lying $^{3}T_{1}$ level with almost the same
parameters ( only two semi-empirical
values differs less than 8 percent). \\
The results of our model have justified the Coulson and Kearsly
semi-empirical parameters and also their guess$^{33}$ which
expected that the $t=-8$ eV value be more realistic than
$t=-16.34$ eV . Both values of this parameter have been used in
their models.$^{33,41}$\\
All of these qualitative and quantitatively satisfactory results
in our model have obtained without configuration interaction. The
good quantitative results for $GR1$ and even other reasonable
qualitative results in the Coulson and Kearsly model and all of
the previous molecular approaches $^{22-26,32}$ can only be
achieved with assistance of the Configuration interaction
technique. This is for the first time that a molecular approach
put aside the configuration interaction technique and could
predict more experimental data. Configuration interaction has been
widely used in all of the previous theoretical works to overcome
the disagreement between theoretical results and experiment even
at the qualitative levels.$^{22-26,33,41,42}$ Although precise
review of the previous exact models $^{33,41}$ results on the
charged vacancy with parameter $t=-16.34$ eV shows
that applying configuration interaction gives wrong ground state for $ V^{-}$.\\
Similar ambiguity also exists in the electronic configuration of
the ground and excited state of the GR1. While the models assume
that the ground and excited states of $GR1$ arise from the single
configuration $a^{2}t^{2}$, at the same time they try to find or
use some semi-empirically coefficients for the contribution of the
other possible configurations in the ground and excited states of
$GR1$.$^{21,26}$ Here we refer to Davies $^{4}$ remark in this
regard:\\"The importance of configuration interaction at the
vacancy in diamond was controversial, and alternative approaches
were investigated in which emphasize was on the benefit of using
large cluster".\\The main advantages of the presented formalism
which is based on molecular model is removing the configuration
interaction besides it's satisfactory quantitative results. By
this means our model have solved one important shortcoming of the
molecular approaches with respect to cluster models or density
functional approaches.\\ In the present model, the resultant
eigenstates of the Hamiltonian have unique expansion on the
starting constructed basis in configuration space, so
configuration interaction comes out in a natural manner from the
formalism. From this point, it is possible to find the
contribution of each electronic configuration in ground and
excited states of the system.\\ For $GR1$ transition, the results
of calculation for the ground state $^{1}E$ and the excited state
$^{1}T_{2}$ with the justified parameters are illustrated in Fig.
5. The possible electronic configuration for states with value of
$S=0$ are: $(1,1,1,1), (2,1,1,0), (2,2,0,0)$ where the numbers in
parenthesis show the occupation number of each tetrahedral orbital
of vacancy system. As it is shown in Fig. 5,
\begin{center}
\begin{figure}[t]
\includegraphics[width=5cm,height=7cm,angle=270]{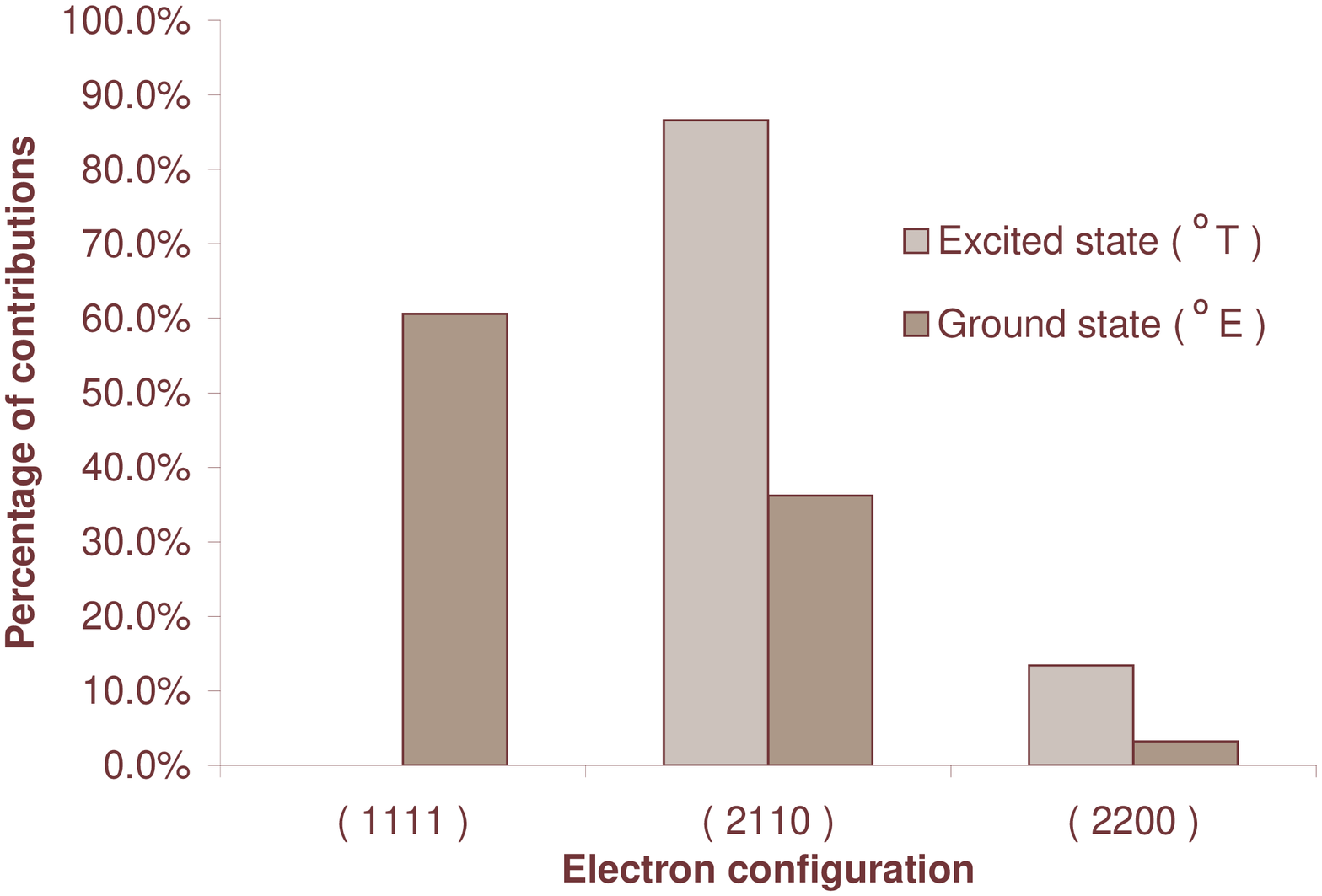}
\caption{ Percentage of the contribution of each possible
electronic configuration in ground and excited state of GR1
transition for neutral vacancy system, the numbers in parenthesis
are occupation number of each tetrahedral orbital}
\end{figure}
\end{center}
all possible electronic configurations are present in the ground
state $^{1}E$. The most probable configuration in ground state
arise from $(1,1,1,1)$ configuration ($61$ $ percent $) which has
only singly occupied orbitals. Two doubly occupied configurations
have very low contribution (3 percent) in the ground state. In
$(1,1,1,1)$ configuration the electrons have maximum separation
distance, which results in lowering the Coulombic repulsion energy
of the system. In contrast to this, the excited state of $GR1$,
i.e. $^{1}T_{2}$ is orthogonal to $(1,1,1,1)$ configuration and is
almost a purely doubly occupied configuration state. This
configuration has the highest Coulombic repulsion energy due to
its lowest separation distance between electrons of the
system.\\Roughly speaking, we can state that $GR1$ transition is
resultant of a transition from higher probability configuration $(1,1,1,1)$ and 
$(2,1,1,0)$ configurations of the system to $(2,1,1,0)$ configuration .\\
Our results modifies the common belief about ground and excited
states of the $GR1$ transition that assumes they both
belong to $a^{2}t^{2}$ configuration.$^{21,26,33,41,42,44,45}$ \\
As the results of our calculation show, the absence of $(1,1,1,1)$
configuration in the ground state of the system is unexpected,
since it has minimum Coulombic repulsion energy among the other
configurations of the system. Also it is the only compatible
configuration of the unique $^{5}A_{2}$ excited state, which in
the delocalization limit ($t>-5.96$ eV) is the ground state of the
system.\\ For $ND1$ transition, since the ground and exited states
have the spin equal to $3/2$ the only configuration which can be
encountered is $(2,1,1,1)$. Therefore the ground and excited
states are only a combination of these configurations with
different $S_{z}$ values and the results are same as other models .\\
The new information about the ground and excited state of the
$GR1$ is very important, It can open a new opportunity for
ab-initio density functional approaches which have not been
attempted to estimate the value of this transition.$^{35}$ This is
due to the results of the previous molecular orbital models which
have assumed the ground and excited state belong to the same
configuration $a^{2}t^{2}$. However by the new result of present
model, they can try to obtain the transition energy of the $GR1$
transition similar to $ND1$. These information are also useful for
theoretical approaches which start from a guess wave function for
ground and excited states of such a molecular system.
\section{Conclusion}
Theoretical studies of the vacancy electronic states in diamond
have been a challenging problem for the five decades. Localized
models which originate from Coulson and Kearsly pioneering work,
predict correctly the spatial symmetry and spin of electronic
states but can not give quantitatively good results for electronic
transitions in diamond vacancies. In their model despite of
excellent theoretical framework and strong physical intuition, the
computational method are very old and can not be adapted with
nowadays powerful computer assisted methods. \\ We have developed
a new approach based on generalized Hubbard Hamiltonian and many
body techniques. A complete set of atomic orbital basis for
describing this system is constructed according to $S_{z}$
representation of each electronic configuration in the four
vacancy orbital. This confines the calculation only in definite
$S_{z}$ subspace. This representation can simply estimate number
of states of the system and also each definite $S_{z}$ block size
of Hamiltonian. This is in contrast to previous models that uses
molecular orbital bases and manually construct each
symmetry and spin adapted wave functions. \\
The generalized Hubbard Hamiltonian which contains whole of the
$e-e$ correlation terms is included with spatial symmetry
information of the vacancy therefore the number of the independent
parameters of the Hamiltonian is reduced to $8$. By this means the
effect of the parameters on the energy states directly can be
examined. A detailed computer software is developed to solve this
Hamiltonian . The resultant Hamiltonian matrix is transformed to
change basis from $S_{z}$ to $S^{2}$ representation. As the
result, the eigenstates have definite and correct spin and
symmetry properties. \\ The effect of changing each Hamiltonian
parameters on the electronic states of the system directly can be
examined. This calculation shows that for obtaining experimentally
correct results for the ground state of $V^{0}$ and $V^{-}$,
parameter $t$ should be in the $-10.22$ eV to $-6.95$ eV range,
although increasing $t$ in this range has low effect on $GR1$ but
significantly increases $ND1$ transition energy. According to this
model the optimum value for two key semi-empirical parameters $t$
(in allowed range) and $U$ are $-7.747$ eV, $13.58$ eV
respectively. Two famous absorption lines of $V^{0}$ and $V^{-}$
namely $GR1$ ($1.67$ eV) and $ND1$ ($3.15$ eV) can be obtained
simultaneously and exactly with these parameters in addition to
the location of the low-lying $^{3}T_{1}$ state at $113$ meV above
the ground state of the $V^{0}$. These results have been obtained
for the time by such justifiable parameters. These parameters also
justify already reported semi-empirical values that only could
estimate $GR1$ transition and gave experimentally wrong results
for $^{3}T_{1}$.
\\For the first time the configuration interaction procedure,
has not used in such a calculations and it arises in a natural
from from the formalism. Despite of its ambiguous physical nature,
the configuration interaction techniques plays a fundamental and
vital role in reaching to an agreement with experimental data in
previous localized models. This vital role even exist at the level
of qualitatively acceptable results. This point enables us to
estimate the contribution of each electronic configuration in
ground and excited states of the system. The results for $GR1$
absorption line show that it is almost a transition from
configurations that consist of a pure singly occupied orbital and
a singly double occupied orbital to a singly occupied orbital
configuration. This new information can be used by the density
functional approache to this problem. The results of our model
justifies the first suggested molecular model by the Coulson and
Kearsly is the five decades ago and shows that their model has
enough physical depth to accompany with a new computational
methods to obtain very satisfactory quantitative results. The
results also show the benefit of the localized models with respect
to cluster or density functional approaches by no need of
configuration interaction .
\\
\section{Acknowledgment}
We would like to thanks Dr.A. Shafikhani and Mr.M. Saiari for
their valuable comments about this work. We acknowledge the help
of Mr.A. Rezakhani(Phd student) and Mr.M. Ghasempour for computer
programming and paper preparation.
\\\\\\ $^{*}$e-mail: Heydaris@mehr.sharif.edu\\
    $^{\dag}$e-mail: Vesaghi@sharif.edu
\\ $^{\ddag}$e-mail: K1@sharif.edu

\end{document}